\documentclass[10pt, twocolumn, letterpaper]{article}

\usepackage[utf8]{inputenc}
\usepackage[T1]{fontenc}
\usepackage{cuted}
\usepackage{algorithm}
\usepackage{algpseudocode}
\usepackage{enumerate}
\usepackage{newtxtext, newtxmath} 
\usepackage{geometry}
\usepackage{multirow}
\usepackage{makecell}
\usepackage{xcolor}
\usepackage{graphicx}
\usepackage{fancyhdr}
\usepackage{lipsum} 
\usepackage{tikz}   
\usepackage{titlesec} 
\usepackage{booktabs} 
\usepackage{authblk}  
\usepackage[hidelinks, colorlinks=true]{hyperref} 
\usepackage{eso-pic}  
\usepackage{dblfloatfix}

\usepackage[most]{tcolorbox}

\definecolor{OkestroBlue}{HTML}{0033A0} 
\definecolor{OkestroDark}{HTML}{001540} 
\definecolor{OkestroGrey}{HTML}{777777}
\definecolor{AbstractBG}{HTML}{E9E9E9}

\hypersetup{
    linkcolor=OkestroBlue,
    citecolor=OkestroBlue,
    urlcolor=OkestroBlue,
    pdftitle={Okestro Research Paper}
}

\AddToShipoutPictureBG*{%
    \begin{tikzpicture}[remember picture, overlay]
        \node[opacity=0.04] at (current page.south east) {
            \begin{tikzpicture}
                \fill[OkestroBlue] (-2,2) circle (12cm);
            \end{tikzpicture}
        };
        \node[opacity=0.03] at (current page.east) {
            \begin{tikzpicture}
                \fill[OkestroBlue] (0,5) circle (8cm);
            \end{tikzpicture}
        };
    \end{tikzpicture}
}

\titleformat{\section}
  {\sffamily\large\bfseries}
  {\thesection}{1em}{}
\titleformat{\subsection}
  {\sffamily\large\bfseries}
  {\thesubsection}{1em}{}

\makeatletter
\def\@tablestring{table}
\let\@oldmakecaption\@makecaption
\long\def\@makecaption#1#2{%
  \ifx\@captype\@tablestring
    \vskip\abovecaptionskip
    \centering #1: #2\par
    \vskip\belowcaptionskip
  \else
    \@oldmakecaption{#1}{#2}%
  \fi}
\makeatother

\newtcolorbox{abstractbox}{
  enhanced,
  colback=white,
  colframe=AbstractBG,
  arc=6pt,
  boxrule=0.6pt,
  left=8pt,
  right=8pt,
  top=8pt,
  bottom=8pt,
  width=0.88\textwidth
}

\title{
  \sffamily\bfseries
  Quantifying Autoscaler Vulnerabilities: An Empirical Study of Resource Misallocation Induced by Cloud Infrastructure Faults
}

\author{
  {\large
    Gijun Park\textsuperscript{1,*}
  }\\
  {\small
    \textsuperscript{1}Okestro AI Research Center, Seoul, Republic of Korea
  }
}

\makeatletter

\footnotetext[1]{Corresponding author: Gijun Park (gj.park@okestro.com)}

\makeatother

\date{}

\begin{document}
\begin{strip}

\vspace{-5.0em}

\maketitle

\thispagestyle{fancy}

\begin{center}

\begin{abstractbox}
\small
\textbf{Abstract.}
Resource autoscaling mechanisms in cloud environments depend on accurate performance metrics to make optimal provisioning decisions. When infrastructure faults—including hardware malfunctions, network disruptions, and software anomalies—corrupt these metrics, autoscalers may systematically over- or under-provision resources, resulting in elevated operational expenses or degraded service reliability. This paper conducts controlled simulation experiments to measure how four prevalent fault categories affect both vertical and horizontal autoscaling behaviors across multiple instance configurations and service level objective (SLO) thresholds. Experimental findings demonstrate that storage-related faults generate the largest cost overhead, adding up to \$258 monthly under horizontal scaling policies, whereas routing anomalies consistently bias autoscalers toward insufficient resource allocation. The sensitivity to fault-induced metric distortions differs markedly between scaling strategies: horizontal autoscaling exhibits greater susceptibility to transient anomalies, particularly near threshold boundaries. These empirically-grounded insights offer actionable recommendations for designing fault-tolerant autoscaling policies that distinguish genuine workload fluctuations from failure artifacts.

\vspace{1.2em}

\end{abstractbox}
\end{center}
\vspace{0.5cm}

\end{strip}

\section{Introduction}
In the field of cloud computing, efficient and stable resource management has emerged as a key challenge for sustainable operations. \cite{aslam_information_2017,mustafa_resource_2015} While cloud systems are characterized by elastic resource scaling to accommodate fluctuating workloads, the pay-as-you-go billing model can occasionally result in higher costs for users who consistently consume extensive resources than an equivalent on-premises infrastructure. \cite{mireslami_dynamic_2021}

Cloud autoscalers are designed to automatically scale resources such as virtual machines (VMs) or containers to maintain optimal performance while minimizing resource waste, with intelligent algorithms based on monitoring of performance metrics such as resource usage or request latency. \cite{tamiru_experimental_2020} This dynamic resource allocation enables systems to maintain objective performance and availability under variable workload conditions. \cite{endo_high_2016} However, the reliability and effectiveness of autoscalers is highly dependent on the accuracy of performance metrics, which can be significantly impacted by cloud failures. \cite{taherizadeh_key_2020}

Cloud failures, such as hardware malfunctions, network outages, or software problems, can introduce outliers in performance metrics that result in incorrect autoscaler behavior. \cite{herbst_quantifying_2018,xu_towards_2020,kalavri_three_2018} When performance metrics cannot accurately represent the actual workload size due to symptoms that failures cause and derive lower values, the autoscaler may operate in an underscaling. In contrast, higher performance metric values that are associated with failure symptoms can lead to overscaling. These scenarios can lead to critical issues, such as service downtime due to insufficient resources or additional unnecessary costs resulting from overprovisioning.

While cloud failures are widely presumed to degrade performance, including resource exhaustion, increased request latency, reduced throughput, and inflate costs, few studies have provided precise and quantitative insights into how specific failures affect resource management. This study addresses the following key research questions.

\begin{enumerate}[1.]
\item \textbf{RQ1}: How do distinct, common cloud failures (e.g., disk, software, network faults) alter CPU usage and other resource metrics that autoscalers rely on?
\item \textbf{RQ2}: How much cost overhead or reliability risk arises from these altered metrics under both vertical and horizontal scaling across different instance families?
\item \textbf{RQ3}: How do different thresholds to triggering the scaling such as service level objective (SLO) of 50\% vs 85\% resource utilization affect the magnitude and cost impact of failure-induced resource misallocation?
\end{enumerate}

In this study, the impact of various failures that commonly occur in cloud systems on autoscaler behavior and cloud usage cost are evaluated by scientific experimentation and verified. Specifically, the difference between each failure symptom and a normal state in terms of performance metrics is analyzed. In addition, the effects of different types of failures on certain elements of both vertical and horizontal autoscaling are assessed.

\section{Related Work}

Ravichandiran et al. proposed an anomaly detection mechanism that uses resource behavior analysis to prevent economic denial of sustainability (EDoS) and resource wastage in autoscaling systems in cloud environments. \cite{ravichandiran_anomaly_2018} Moghaddam et al. proposed a two-level autoscaling framework that combines anomaly detection with horizontal and vertical scaling strategies to address performance issues in cloud environments. \cite{moghaddam_acas_2019} The framework uses an unsupervised isolation forest-based anomaly detection method to analyze VM performance metrics and predict future anomalies, enabling proactive scaling decisions.

Lalropuia et al. proposed a state-based availability model using a semi-Markovian process to assess steady-state availability of the cloud under EDoS attacks. \cite{lalropuia_availability_2021} Ahmed et al. proposed an experimental framework with application-level fault injection (ALFI) to study the effect of faults on the scalability behavior of cloud services. \cite{al-said_ahmad_scalability_2022} They simulated delay latency injection at two different times using a real-world cloud-based software service in the EC2 cloud. They then compared the results with baseline data to investigate the scalability resilience of the cloud-based software service. Kesavan et al. investigated fault-scalable virtualized infrastructure management to improve the resilience of infrastructure-as-a-service (IaaS) management stacks. \cite{kesavan_fault-scalable_2017} They proposed an autoscaler with VM speculative replication that increases the likelihood that at least one copy will succeed, even if some attempts fail due to faults, by simultaneously executing multiple copies on different physical resources as extensions to the IaaS stacks.

While previous research has focused predominantly on improving proactive failure detection and resilience of infrastructure or services, relatively little is known about systematically quantifying how these disruptions affect autoscaling outcomes and costs. Existing studies that acknowledge the economic impact of failures typically examine only specific fault types or rely on partial workloads that do not capture broader operational realities. Moreover, they have rarely provided side-by-side comparisons of multiple disruptions under varying SLO thresholds or across different instance types, which makes it difficult to understand how certain failures shape resource decisions and overall expenditures and how the impacts between failures differ relatively.

\section{Fualt-aware Autoscaling}

Autoscaling algorithms typically evaluate the appropriateness of sizing and resizing resources based on performance metrics, such as CPU utilization or requests per second (RPS), over a historical observation period. However, if the data that trigger this autoscaling process represent inaccurate values due to failures, the autoscaling algorithm may incorrectly consider the resources to be underscaled or overscaled, regardless of the actual workload state.

The major failures that can occur in cloud systems include network failures, hardware failures, software problems, scalability and load problems, security breaches, and environmental failures. \cite{yousif_cloud_2018,garraghan_emergent_2018,wang_approaches_2018,liu_early_2024,tengku_asmawi_cloud_2022} Network failures include faulty routing, which disrupts data transmission and can cause service unavailability. Hardware failures encompass CPU overload, memory exhaustion, and I/O bottlenecks that degrade the processing capacity and performance of the overall system. In terms of software problems, such as implementation bugs which are coding defects that cause unexpected behavior. Security breaches include denial-of-service (DoS) attacks that overwhelm system resources and malware infections that compromise system integrity and can lead to data corruption or unauthorized access. In this study, we focus on common and important failures among a wide variety of failures in a cloud system to enhance the practical applicability of the experimental results. These failures are selected by identifying them based on their impact on autoscalers in terms of impact scope, detection characteristics, and business relevance.

Network failures, from the perspective of scope of impact, propagate across multiple layers due to system interconnectivity: at the infrastructure level, they affect routing and physical connections; at the application level, they hinder API calls and microservice interactions. \cite{mathews_insights_2021} Hardware failures can also cascade across these layers, affecting processing efficiency and causing broader system slowdowns. \cite{chen_container_2023,dong_understanding_2025} Software problems cause disruption of the application functionality. Security breaches like DoS attacks typically start at the infrastructure level, overwhelming network resources, and then impacting platform services and application performance. \cite{agrawal_defense_2019} From the perspective of detection characteristics, network failures can lead to increased latency and packet loss, which an autoscaler could misinterpret as a load surge. \cite{wang_mining_2023} Similarly, hardware failures can create the false impression of increased workload because they present symptoms of higher resource utilization. \cite{strikos_breaking_2024} Software problems can cause resource leaks or sudden crashes, which can cause the autoscaler to misdeploy resources unnecessarily. \cite{8891707} In terms of business relevance, hardware failures and software problems demonstrate high recovery complexity and require extensive resources and time to resolve, while network failures directly affect service availability metrics that are crucial to maintaining service level agreements (SLAs). \cite{bauer_hardware_2012,mo_efficient_2022}

Consequently, this study focuses on four types of failures: network failures, hardware failures, software problems, and DoS attacks. These failures are critical considerations for site reliability engineering because they interfere with the normal operation of the autoscaler by leading to abnormal performance metrics, as well as accounting for a significant portion of the root causes of incident reports. \cite{fiondella_cloud_2013}

\subsection{Fault Setup}

As both hardware and network failures, router failures occur when hardware malfunctions or network virtualization-related anomalies within the routing infrastructure induce abnormal communication delays. In the same class, disk bottleneck occurs when block storage mounted on a virtual machine fails, preventing normal disk I/O request processing. This failure is created by temporarily disconnecting the block storage from the VMs. Upon reconnection, the accumulated I/O operations create a bottleneck as pending requests are processed.

Software problems encompass various failures ranging from intermittent request failures to service outages during application development, integration, deployment, and operations. This study reproduces request failures caused by API implementation defects. A typical DoS attack in which abnormal traffic is sent to a target, depleting the target's computing resources, and causing disruptions in normal services. \cite{yan_distributed_2015,salim_distributed_2020} The attacks examined include volumetric attacks that target Layer 7 (L7) and synchronize sequence number (SYN) flooding and UDP flooding attacks that target Layer 4 (L4).

\subsection{Autoscaler}

Autoscaling works the following way: if the performance metrics declared by triggers exceed the optimal level, it is deemed underprovisioning, and additional computing resources are allocated to improve the operational stability of the application. In contrast, if the performance metric falls below the optimal level, it is considered overprovisioning, prompting the reduction of unnecessary computing resources to reduce costs. Autoscaling can be broadly categorized into two approaches. The method of managing the load by directly adding or subtracting the computing resources of VMs is called vertical autoscaling, whereas the approach of managing the load of individual VMs by increasing the number of replicas to process workloads in parallel is called horizontal autoscaling.

The simulation was performed in both vertical and horizontal autoscaling scenarios with the Kubernetes default optimization algorithm, the container orchestration framework commonly used in cloud-native architectures. \cite{noauthor_horizontal_nodate,noauthor_autoscalervertical-pod-autoscaler_nodate} The optimal resource size in the vertical method is derived as in (1), where $i$ is a type of resources, denoted $i \in \{\text{CPU},\text{Memory},\text{DiskIO},\text{NetworkBandwidth}\}$, $m_i$ is the time series array per minute of usage of $i$, $SLO$ is defined as the optimal level of the performance metrics, $\text{spec}_i$ is the current VMs specification for $i$, and $\text{optSpec}_i$ is the optimal VMs specification for $i$.

\begin{equation}
  \text{optSpec}_i= \lceil \text{spec}_i \times (\text{max}(m_i)-(SLO-\text{max}(m_i))) \rceil
\end{equation}

In this context, while the SLO is represented by a single value, performance metrics comprise sequential values recorded at regular intervals throughout the observation period. The SLO is compared with a representative value of performance metrics, derived by maximum aggregation. \cite{qiu_aware_2023} The optimal instance type is determined by grid searching the instance type with all computing specifications that exceeded the appropriate resource size and the lowest cost per hour among the specifications of all available instance types, which is a provisional condition of the cloud service provider (CSP).

The horizontal method uses the same performance metrics as the vertical method to determine whether and to what extent to autoscale; however, the calculation differs. The optimal number of replicas for autoscaling is determined by the product of the current number of replicas and the ratio of the performance metric targets to the representative value of the performance metric. If the representative value of the performance metric exceeds or falls below the targets, replicas are created or deleted to bring it closer to the targets. For example, if the representative value exceeds the target, the replicas increase until it reaches the optimal number of replicas for the representative value. In this case, the optimal number of replicas is derived as shown in (2).

\begin{equation}
  \text{optReplicas}=\text{max}_i( \lceil \text{currentReplicas} \times (\frac{\text{max}(m_i)}{SLO}) \rceil )
\end{equation}

\section{Experiments}

The experimental environment comprises two main components: a server group and a client group. The server group consists of six VMs, which in turn are divided into two subgroups: the experimental group and the control group. The experimental group includes three VMs that are exposed to normal workloads and experience failures. The control group includes three VMs that take only normal workloads. All VMs are implemented using amazon web service (AWS) EC2. The client group contains one VM that generates and transmits workloads to the server group and two VMs that generate DoS attacks. They were created using the Compute Engine service of the Google Cloud Platform. The VMs in the server and client groups are physically located in different data centers to ensure that the experiment reflects the impact of the network in a typical production environment. The autoscaler employs a logical implementation that presupposes immediate reconfiguration of server resources to optimal specifications, as determined by Equations 1 and 2. Mission-critical systems tend to implement more conservative measures to mitigate operational risk, including consideration of the impact of failures on the autoscaler. \cite{dong_understanding_2025} Therefore, it reflects scenarios in which resource utilization is set at 85\% for SLOs that prioritize cost efficiency for services less critical or used infrequently and at 50\% for SLOs that emphasize stability. The general configuration of the experimental environment and the detailed specifications of the VMs in each group are illustrated in Figure 1 and Table 1, respectively.

In the experiment, each VM in the server group is subjected to a normal workload for 12.5 min, followed by the 5 min failure scenario described in Section III with the normal workload. Then, the system is subjected to only the normal workload for 2.5 min. In hands-on practice, the lookback time for deciding on an autoscaling action is usually set to 5 min for an agile response. \cite{noauthor_best_nodate} However, in this study, it was set to 15 min to take into account the different effects of normal operation, fault start, and fault end. Performance metrics data for each VM in the server group are collected via Prometheus, with an interval of 1 s. \cite{noauthor_parsa-epflcloudsuite_2024} The normal workload is generated using the web serving benchmark service in CloudSuite 4.0. \cite{noauthor_prometheusprometheus_2024} The web service benchmark equally reproduces the general functions provided by real-world social media platforms and the actions taken by users (logging in, posting, adding and removing friends, sending messages, etc.) through the internal server module and the client module. It comprises server-side modules that handle service functions, such as the web server, the DB server, and Memcached, and client modules that generate requests for these service functions. These server-side and client modules are deployed as Docker containers on VMs in the server and client groups, respectively.

\begin{figure}[ht]
    \centering
    \includegraphics[width=\columnwidth]{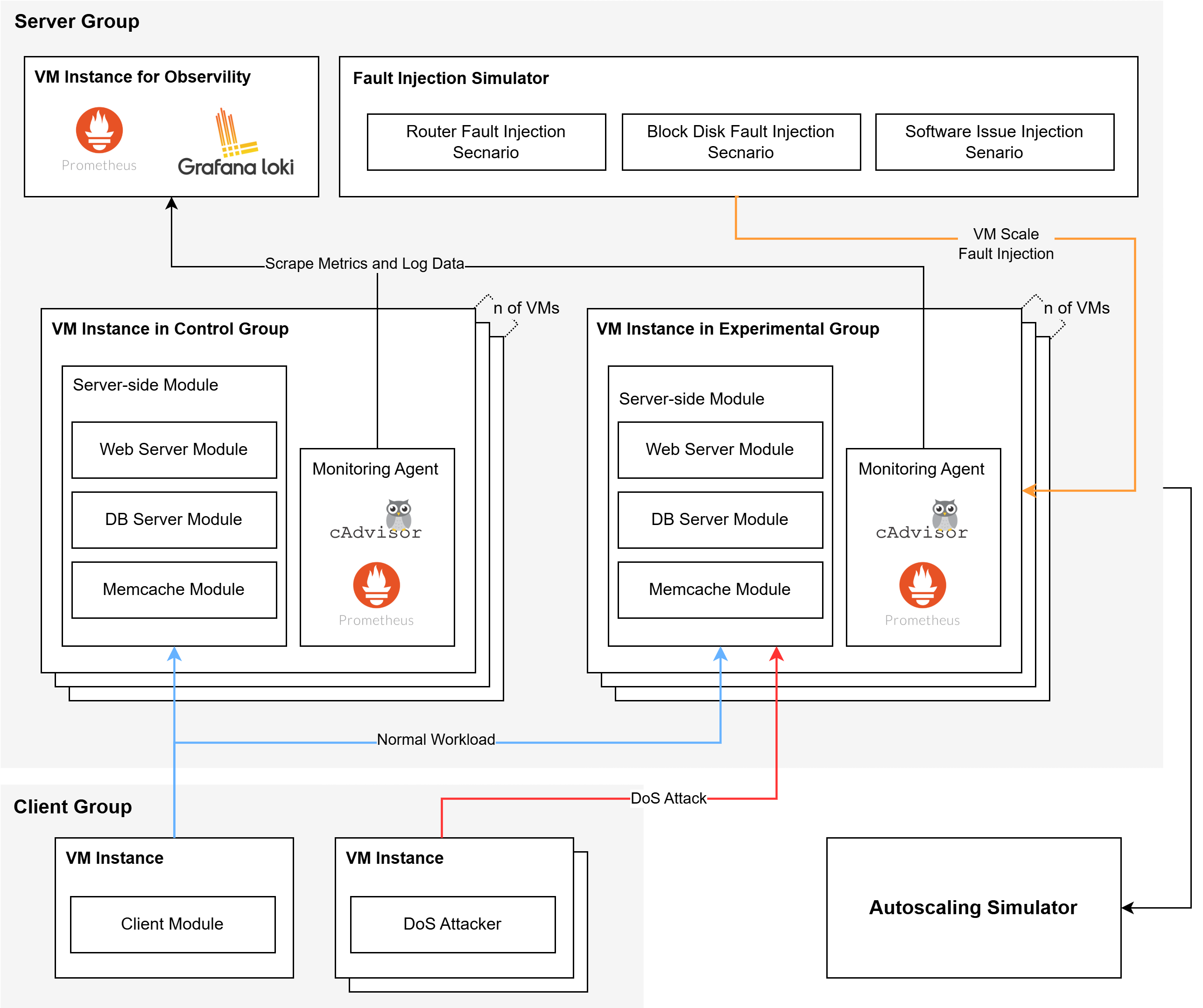}
    \caption{Architecture of the experimental environment, which consists of a server group that includes the experimental group and control group, a client group, and the simulation tools.}
    \label{fig}
\end{figure}

\begingroup
\setlength{\tabcolsep}{3pt}
\begin{table}[h]
    \caption{Detailed Specification of Experimental VM Instance sizes.}
    \vspace{0.1cm}
    \begin{tabular}{ccccccc}
    \toprule
    \multicolumn{2}{c}{\multirow{2}{*}{Instance Type}} & \multirow{3}{*}{\makecell{CPU\\Perf.\\(GHz)}} & \multirow{3}{*}{vCPU} & \multirow{3}{*}
    {\makecell{Memory\\(GB)}} & \multirow{3}{*}{\makecell{Network\\(Gbps)}} & \multirow{3}{*}{\makecell{Cost\\(\$/h)}} \\
    \multicolumn{2}{c}{} &&&&& \\
    Families & Size & & & & & \\
    \midrule
    \multirow{3}{*}{m5} & large & \multirow{6}{*}{3.1} & 2 & 8 & \multirow{3}{*}{Max 5} & 0.104 \\
    & xlarge & & 4 & 16 & & 0.208 \\
    & 2xlarge & & 8 & 32 & & 0.416 \\
    \multirow{3}{*}{t3} & large & & 2 & 8 & \multirow{6}{*}{Max 10} & 0.118 \\
    & xlarge & & 4 & 16 & & 0.236 \\
    & 2xlarge & & 8 & 32 & & 0.482 \\
    \multirow{3}{*}{c5} & large & \multirow{3}{*}{3.3} & 2 & 4 & & 0.086 \\
    & xlarge & & 4 & 8 & & 0.172 \\
    & 2xlarge & & 8 & 16 & & 0.344 \\
    \bottomrule
    \end{tabular}
\end{table}
\endgroup

A normal workload is simulated 5 virtual users who use the service for 15 min. In the case of a DoS attack, the SYN and UDP flooding is generated by the hping3 tool, and the volumetric attacks are performed by the MHDDoS tool. \cite{sanfilippo_antirezhping_2025,noauthor_matrixtmmhddos_2024} The hping3 attack uses a window size of 64 bytes and a data size of 120 bytes, and the MHDDoS attack uses a thread count of 450 and a request rate of 150 requests per second. The attack parameters were determined through an empirical study of values that demonstrated symptoms that included a higher probability of triggering aberrant autoscaling than 50\% compared to normal state operations and actual service interruptions when failures were injected into the given environment. Router failures are simulated by adding 200 millisecond latency and jitter to each VM network interface for all incoming traffic using the AWS Fault Injection Simulator (FIS) tool. \cite{noauthor_best_nodate} This injected latency is set to the default value of the commonly accepted TCP Read Timeout. Disk failures are simulated by performing a volume IO pause scenario that temporarily blocks IO to OS volumes on VMs with the AWS FIS tool. Software problems are generated by causing random packet loss using Pumba, the Docker chaos testing tool in the communication between the web server module and the DB server module of the web serving. \cite{ledenev_alexei-ledpumba_2024} Minor errors or bugs are mitigated by the fault-tolerant design of libraries and frameworks that implement business logic so that symptoms of failure are not readily visible in performance metrics. Therefore, the software problem assumes a catastrophic failure event with a packet loss rate 50\%.

\section{Result And Discussion}

\subsection{Impact of Failures on Vertical Autoscaling}

The case of SYN, UDP flooding attacks, and \textit{large} size instances demonstrate error ratios that remain close to zero, in the 85\% SLO configuration as described in Figure 2 (a). The error ratio quantifies the deviation in resource allocation decisions between failure and normal states, calculated as

\begin{equation}
    E_r = \frac{V_{\text{abnormal}} - V_{\text{normal}}}{V_{\text{normal}}} \times 100%
\end{equation}

\begin{figure*}[!t]
    \centering
    \includegraphics[width=\textwidth]{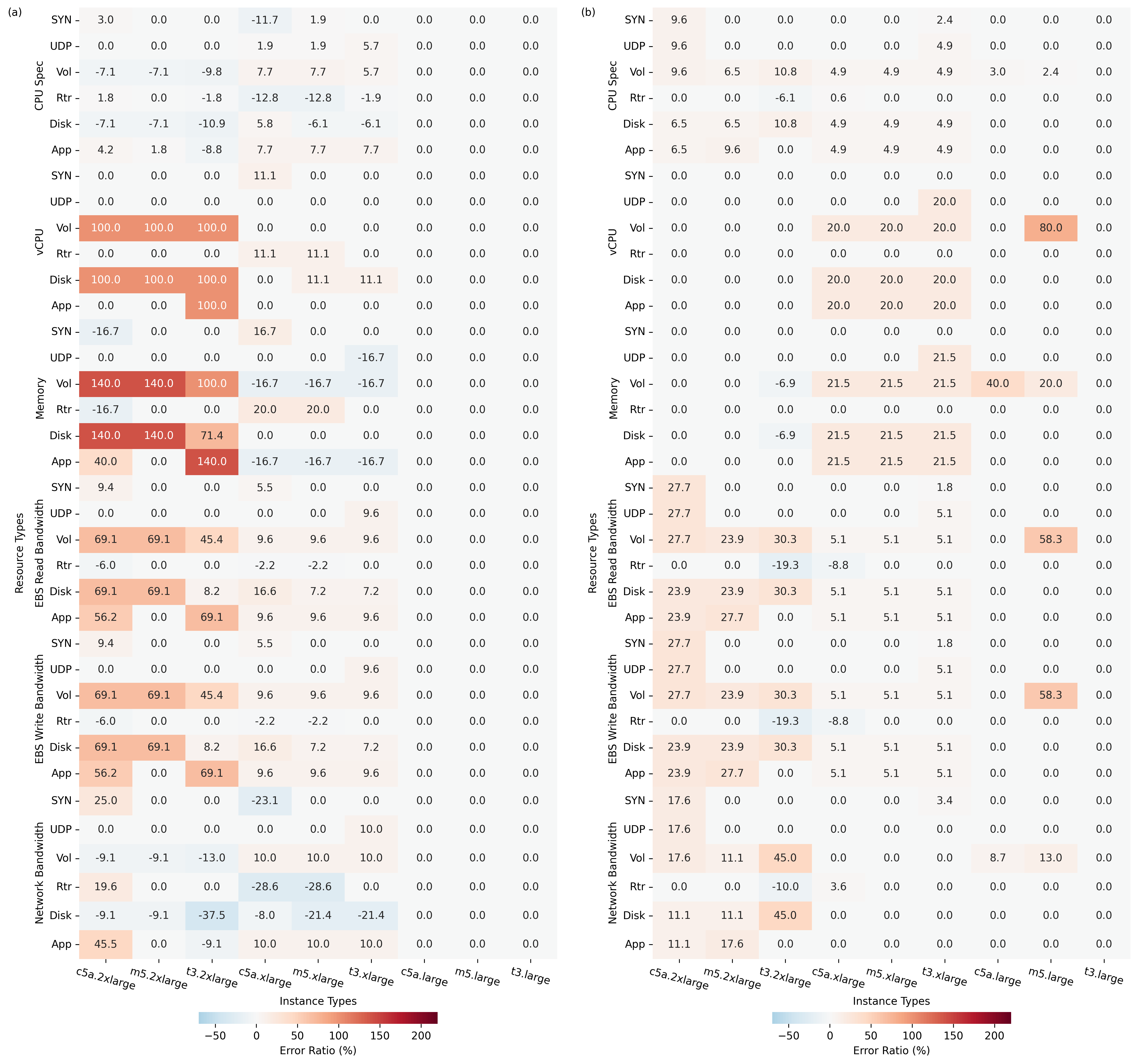}
    \caption{ Comparison of the gap in the optimal specification between normal and failure state by instance size for vertical autoscaling with (a) SLO 85\% and (b) SLO 50\%. Failures include SYN flooding attack (SYN), UDP flooding attack (UDP), volumetric attack (Vol), router failure (Rtr), disk failure (Disk), and software problem (App).}
\end{figure*}

where $E_r$ represents the percentage of the error ratio, $V_{\text{abnormal}}$ denotes the optimal resource specification determined during failure conditions, and $V_{\text{normal}}$ represents the baseline specification under normal operation. Positive error ratios indicate overprovisioning induced by failures, while negative values suggest underprovisioning relative to actual workload requirements. This stability occurs because these attacks saturate CPU utilization to approximately 100\% regardless of the instance size, making the autoscaling decisions consistent between normal and failure states. In contrast, volumetric attacks and disk failures produce substantial positive error ratios, particularly for larger instance types. The c5a.2xlarge configuration shows error ratios that exceeded 140\% for both types of failure. This pattern emerges because these failures cause sustained high CPU utilization that exceeds the actual workload requirements, leading the autoscaler to allocate unnecessarily large instances. Likewise, software problems present overprovisioning, with error ratios of 69.1\% for c5a.2xlarge. This overprovisioning occurs because the temporary increase in CPU usage observed during retry logic and logging procedures triggered by software failures is interpreted as resulting from increased load.

Router failures generally suggest underprovisioning tendencies, with error ratios of -16.7\% for c5a.2xlarge and m5.2xlarge instances. This is due to the network latency caused by the router failure, producing the thread to transition to an I/O waiting state, leading the autoscaler to misinterpret this as a reduced in computational demand.

The SLO scenario 50\% exhibits lesson effects in all types of failure, as shown in Figure 2(b). Volumetric attacks and disk failures continue to cause overprovisioning, but with a decreased error ratio around the level of 20\% and a narrower impact across instance families. In particular, the 2xlarge size shows more faintly the impact of failure under the lower SLO threshold, with error ratios that limit 23.9\% for volumetric attacks in the EBS bandwidth of the m5.2xlarge instance. This decrement in sensitivity occurs because the lower threshold reduces the margin between the normal operation and the scaling trigger point.

\subsection{Impact of Failures on Horizontal Autoscaling}

\begin{figure}[ht]
    \centering
    \includegraphics[width=\columnwidth]{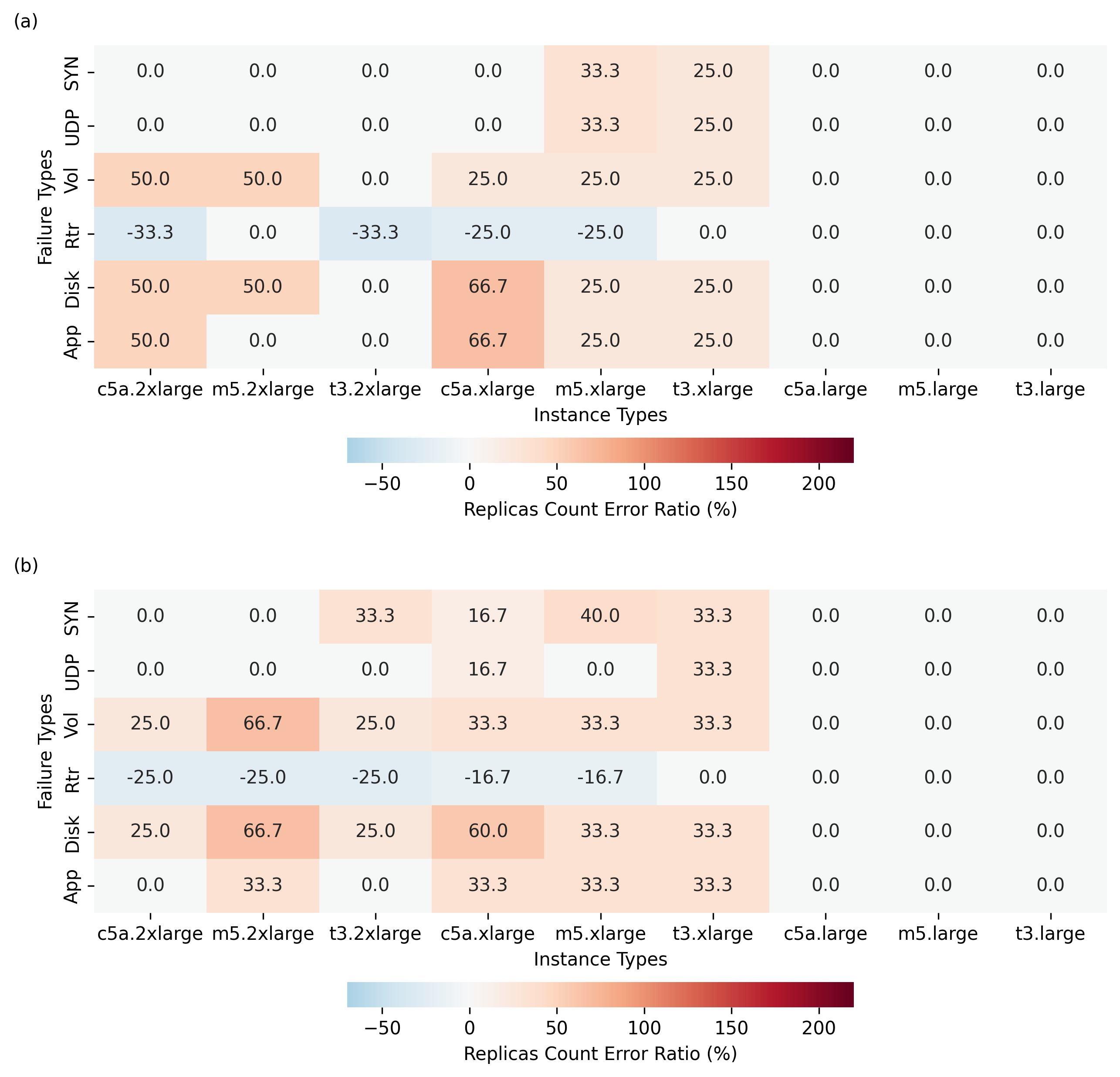}
    \caption{Comparison of the gap in the optimal replicas count between normal and failure state by instance size for horizontal autoscaling with (a) SLO 85\% and (b) SLO 50\%. Failures include SYN flooding attack (SYN), UDP flooding attack (UDP), volumetric attack (Vol), router failure (Rtr), disk failure (Disk), and software problem (App).}
\end{figure}

Horizontal autoscaling shows different scaling patterns than vertical autoscaling despite using identical metrics and thresholds, as shown in Figure 3(a). In 85\% SLO, volumetric attacks cause a 50\% error ratio for c5a.2xlarge, while disk failures produce 66.7\% for c5a.xlarge instances. When disk failures cause CPU utilization to approach 100\%, the autoscaler doubles the replica count, compounding resource misallocation. The t3 family maintains error ratios at or below 33.3\%, indicating that the characteristics of the burstable instance provide a buffer against failure-induced scaling errors.

The SLO scenario 50\% shows an increased sensitivity to failure-induced metric distortions explained in Figure 3 (b). SYN flooding attacks, which had minimal impact in 85\% SLO, now cause 16.7\% overprovisioning for c5a.xlarge instances. This shift occurs because the lower threshold intersects with the failure-induced CPU utilization patterns, triggering unnecessary replica additions. Software problems produce distincted effects depending on the type of instance, with instances of \textit{xlarge} size and m5.2xlarge showing 33.3\% error ratio while others remain unaffected.

The differential impact between vertical and horizontal autoscaling based on their distinct optimization approaches. While vertical autoscaling abstracts performance metrics into instance type decisions, horizontal autoscaling directly translates metric deviations into replica count adjustments. This direct relationship makes horizontal autoscaling more sensitive to transient metric distortions, particularly when operating near threshold boundaries.

\subsection{Economic Impact of Failure-Induced Misallocation}

\begin{figure*}
    \centering
    \includegraphics[width=0.8\textwidth]{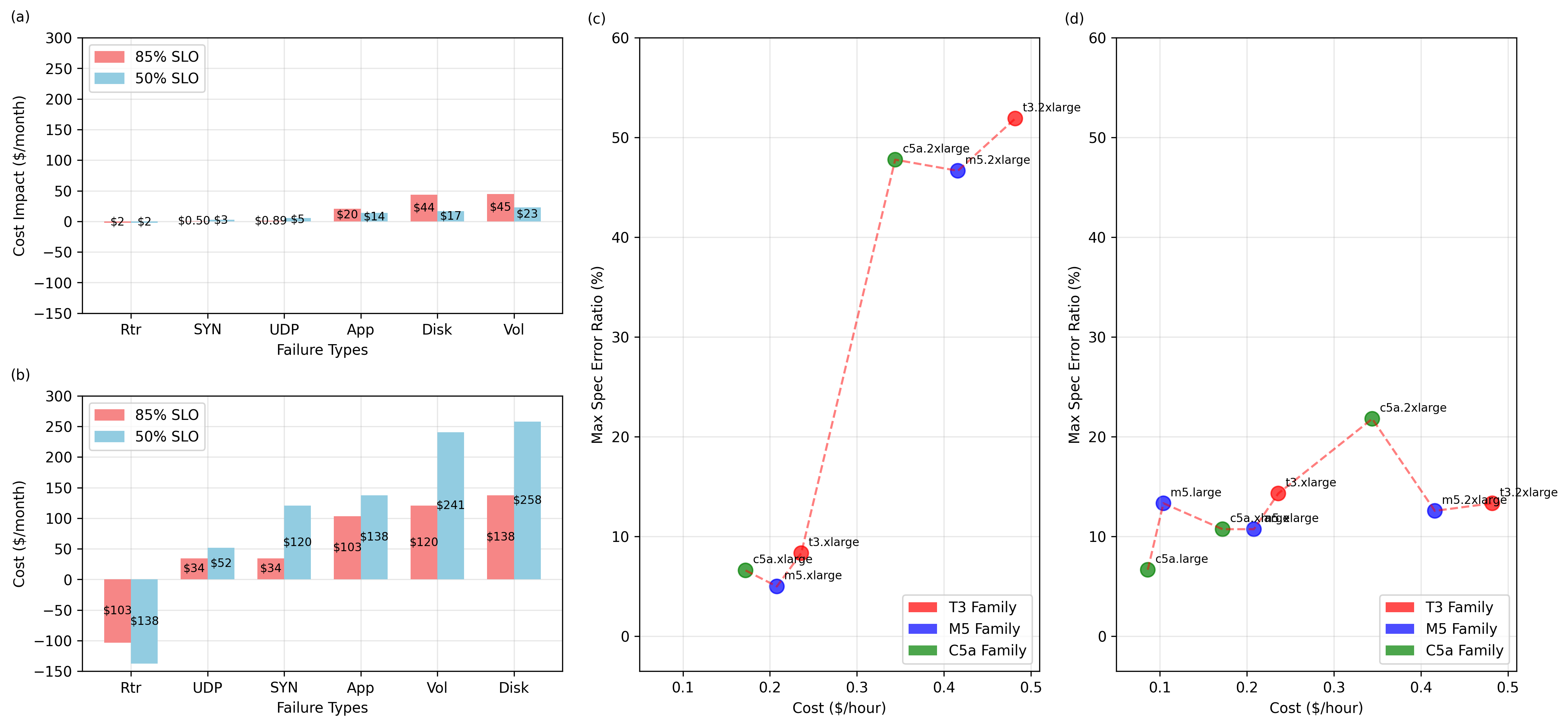}
    \caption{Comparison of the gap in monthly optimal cost between the normal and failure states, by failure type in (a) vertical autoscaling and (b) horizontal autoscaling, comparison of the ratio of  resource insufficient risk caused by the failure and expense, by instance types in (c) SLO 85\% and (d) SLO 50\% vertical autoscaling for total failures.}
\end{figure*}

Quantifying the cost implications of autoscaling errors under various failure conditions reveals, in vertical autoscaling with 85\% SLO, disk failures and volumetric attacks impose the highest economic burden, incurring additional costs of approximately \$44 and \$45 per month, respectively, as shown in Figure 4(a).

Horizontal autoscaling substantially amplifies these economic impacts due to replica multiplication, as described in Figure 4 (b). With 85\% SLO, disk failures incur the highest additional costs at approximately \$258 per month, closely followed by volumetric attacks at \$241 per month. However, router failures in horizontal autoscaling lead to cost reductions of \$103-138 per month depending on the SLO configuration, representing significant underprovisioning that could compromise service availability.

In the 50\% SLO threshold, in vertical autoscaling, cost differences decrease, whereas in horizontal autoscaling, they increase. This is because the lower threshold of 50\% is closed to normal CPU utilization. As a result, smaller metric fluctuations caused by failures are more likely to cross the threshold and trigger scaling actions. Consequently, vertical autoscaling allows for relatively fine-grained control over resource specifications, resulting in smaller cost differences. In contrast, horizontal autoscaling results in larger cost differences because metric fluctuations have a greater cost impact when adding or removing entire replica instances, regardless of threshold sensitivity.

The cost impact varies significantly by instance family and size, as shown in Figures 4(c) and (d). The c5a family, despite offering superior computational performance, consistently incurs higher costs during failure-induced overprovisioning. This relationship suggests a trade-off between performance headroom and economic risk. Smaller instance types generally exhibit lower absolute cost variations but higher relative cost increases, while larger instances show the opposite pattern.

\subsection{Mitigation Strategies and Design Considerations}

The scaling inaccuracies and cost difference identified in our experiments give CSPs an opportunity to concretize vulnerabilities in managed autoscaling service and to quantify the potential impact on service quality and costs for customers. Cloud service users could select a vertical or horizontal scaling methodology suitable for applications, implement failover mitigation strategies, and set SLOs for fault tolerance, based on the observed performance patterns.

More concretely, for disk failure that is characterized by persistent symptoms such as I/O task interruption and CPU utilization approaching 100\%, implementing I/O monitoring that gradually triggers during allowance down time helps to distinguish between increased normal workload and temporary I/O bottlenecks. Particularly, caution should be exercised when radical autoscaling advanced VMs such as compute-optimized or accelerator instances, which are likely to incur excessive costs. For DoS attacks such as SYN and UDP flooding attacks, the fault tolerance of autoscalers can potentially be improved by configuring triggers that activate when multiple complementary metrics less impact by the failure exceed their predefined thresholds simultaneously, such as CPU utilization and network throughput that present negative correlation during the failure, rather than relying solely on default metrics. Router failures, which have shown a tendency to under-provision across instance families could help maintain service reliability for latency-sensitive applications by latency-aware scaling policies that observe network latency thresholds and incorporate them into triggers.

In terms of architectural decisions, organizations that have a margin in error budget on service reliability might purpose the cost effectiveness of scaling with prioritizing vertical autoscaling. While vertical autoscaling has difficulty responding to significant changes in workload and is less responsive than horizontal autoscaling, it shows less variation in behavior between normal and failure states. This consistency helps to prevent unintended autoscaling operations that could be triggered by failures.

\section{Conclusion}

This research presents a comprehensive quantitative evaluation of how common cloud failures impact autoscaling mechanisms through systematic simulation-based experiments. By analyzing the behavior of vertical and horizontal autoscaling in various failure scenarios, we demonstrate that failures introduce significant distortions in performance metrics, leading to resource misallocation with measurable economic consequences.

Our experimental results reveal three findings. First, disk failures impose the most severe economic burden, with additional costs reaching up to \$258 per month in horizontal autoscaling scenarios, while network failures can mislead autoscalers into systematic underprovisioning. Second, the impact of failures varies significantly across instance families and autoscaling strategies, with compute-optimized instances showing greater vulnerability to failure-induced overprovisioning. Third, SLO thresholds fundamentally influence the magnitude of resource misallocation, with lower thresholds exhibiting increased sensitivity to metric distortions but reduced absolute cost impact. These findings establish baseline metrics for integrating failure awareness into autoscaling decision algorithms, enabling autosacler to differentiate between workload-driven and failure-induced resource demands through multi-metric validation and asymmetric scaling thresholds. As a result, both cloud providers and users realize autoscaling strategies that minimize cost-risk trade-offs while improving service continuity under failure conditions.

However, while the CloudSuite 4.0 web serving benchmark that was used as the simulation environment reproduced the same functionality and workloads as a real-world application, the production environment could have suffered high variability in traffic and resource competition compared to the simulation due to the complexity of the overall cloud architecture and various external variables. The fault injection scenario represented major failures in a typical cloud usage environment; however, out-of-cover less frequent but impactful issues could be more important in special environments, such as cascading failures and multi-region outages. In the future, we plan to analyze more compound failures across multiple application types, architectural patterns, and workload characteristics based on the findings of the study and to examine the impact of those failures and the degradation of performance in a high availability configuration, including autoscaling.

\renewcommand{\refname}{\normalsize References}

\begingroup
\small
\bibliographystyle{IEEEtran}
\bibliography{references}
\endgroup

\end{document}